\documentclass[aps,pra,epsfigure,onecolumn,superscriptaddress,notitlepage]{revtex4-1}
\usepackage[colorlinks=true,linkcolor=blue,urlcolor=blue,citecolor=blue,pdfusetitle]{hyperref}
\usepackage[utf8]{inputenc}
\usepackage[english]{babel}
\usepackage{amsmath}
\usepackage[caption = false]{subfig}
\usepackage{graphicx,epstopdf}
\usepackage{blindtext}
\usepackage[table,xcdraw]{xcolor}
\usepackage{lipsum}
\usepackage{amsfonts}
\usepackage{bbm}
\usepackage{amssymb}
\usepackage{enumerate}
\usepackage{color}
\usepackage{latexsym}
\usepackage{times,txfonts}

\newcommand{\Acal}{\mathcal{A}}
\newcommand{\Bcal}{\mathcal{B}}

\newcommand{\Hcal}{\mathcal{H}}

\newcommand{\Lcal}{\mathcal{L}}
\newcommand{\Mcal}{\mathcal{M}}

\newcommand{\Rcal}{\mathcal{R}}

\newcommand{\1}{\mathbbm{1}}
\newcommand{\Lmath}{\mathbbm{L}}
\newcommand{\Rmath}{\mathbbm{R}}

\newcommand{\Lcalb}{\mathfrak{L}}

\newcommand{\ket}[1]{| #1 \rangle}
\newcommand{\bra}[1]{\langle #1 |}
\newcommand{\dket}[1]{| #1 \rangle\rangle}
\newcommand{\dbra}[1]{\langle\langle #1 |}

\newcommand{\dinterpro}[2]{\langle \langle #1 | #2 \rangle \rangle}

\newcommand{\trs}[1]{ \mathrm{tr}[ #1 ]}

\newcommand{\SubFig}[2]{\ref{#1}{\color{blue}#2}}

\definecolor{MyGreen}{RGB}{0, 179, 134}
\definecolor{MyRed}{RGB}{255, 102, 102}

\newcommand{\UFCA}{Centro de Ci\^{e}ncias e Tecnologia, Universidade Federal do Cariri,	63048-080, Juazeiro do Norte, Cear\'{a}, Brazil}

\newcommand{\CSIC}{Instituto de Física Fundamental (IFF), Consejo Superior de Investigaciones Científicas (CSIC), Calle Serrano 113b, 28006 Madrid, Spain}

\usepackage{orcidlink}

\begin{document}

\newcommand{\Title}{Enhanced Quantum Mpemba Effect with Squeezed Thermal Reservoirs}
\title{\Title}

\author{J. Furtado~\orcidlink{0000-0002-1273-519X}}
\email{job.furtado@ufca.edu.br}
\affiliation{\UFCA}
%\affiliation{\Gazi}

\author{Alan C. Santos~\orcidlink{0000-0002-6989-7958}}
\email{ac\_santos@iff.csic.es}
\affiliation{\CSIC}

\begin{abstract}
	The phenomenon where a quantum system can be exponentially accelerated to its stationary state has been referred to as the Quantum Mpemba Effect (QMpE). Due to its analogy with the classical Mpemba effect, \textit{hot water freezes faster than cold water}, this phenomenon has garnered significant attention. Although QMpE has been characterized and experimentally verified in different scenarios, the sufficient and necessary conditions to achieve such a phenomenon are still under investigation. In this paper, we address a sufficient condition for QMpE through a general approach for open quantum system dynamics. With the help of the \textit{Mpemba parameter} introduced in this work to quantify how strong the QMpE can be, we discuss how our conditions can predict and explain the emergence of weak and strong QMpE in a robust way. As an application, by harnessing the intrinsic non-classical nature of squeezed thermal environments, we show how enhanced QMpE can be effectively induced when our conditions are met. We demonstrate that when the system interacts with thermal reservoirs, \textit{a hot qubit freezes faster than a cold qubit} in the presence of squeezing. Our results provide tools and new insights, opening a broad avenue for further investigation at the most fundamental levels of this peculiar phenomenon in the quantum realm.
\end{abstract}

\maketitle

\section{Introduction}

Two physical systems initially at different temperatures, the cold \(T_{\mathrm{c}}\) and hot \(T_{\mathrm{h}} > T_{\mathrm{c}}\), are cooled down by exchanging energy with a colder reservoir at temperature \(T_{0} < T_{\mathrm{c}}\). Even under the same cooling conditions, the hot system can reach a regime of temperatures smaller than those the cold system does during the process. This unconventional effect was observed in the classical physics realm by a 13-year-old high school student, Erasto B. Mpemba, in 1963. The experimental verification of the effect was reported by E. B. Mpemba and D. G. Osborne in 1969~\cite{Mpemba:1969}. F. Carollo \textit{et al}.~\cite{FedericoCarolo:21} introduced a mechanism capable of exponentially speeding up the dynamics of a quantum system to the stationary state, which can be interpreted as Mpemba's effect in the quantum realm. Since then, this quantum Mpemba effect (QMpE) has aroused further investigations about the phenomenon and its characterization in different contexts~\cite{Chatterjee:23,Rylands:23,Kochsiek:22,Ares:23,PhysRevLett.133.140404,Manikandan:21,Rylands:23,Ivander:23,Chatterjee:23-b,Liu:24,Turkeshi:24,Murciano:24,Yamashika:24}.

The QMpE has been investigated in the context of Markovian dynamics of many-body systems through the Dicke model~\cite{FedericoCarolo:21}, the Anderson model (quantum dots reservoirs)~\cite{Chatterjee:23}, the Lieb-Liniger model~\cite{Rylands:23}, and spin chain systems~\cite{Kochsiek:22,Ares:23}, for example. However, to the best of our knowledge, the first experimental evidence of such an effect was provided recently by J. Zhang \textit{et al}.~\cite{Zhang:24} in a single trapped ion system, exploiting the dynamics of a three-level system encoded in the low-lying electronic states of a trapped $^{40}$Ca$^{+}$ ion. However, all these works, among others~\cite{Manikandan:21,Rylands:23,Ivander:23}, have considered the interaction of the physical system with reservoirs described by physical processes other than those observed in thermal reservoirs with a well-defined finite temperature (\(T>0\)). To deal with the proper measure and characterization, some authors~\cite{Chatterjee:23,Chatterjee:23-b} have considered the definition of an alternative version of temperature for isothermal processes, where a detailed study on the exchanged heat and entropy variation of the system is considered.

\begin{figure}
	\centering
	\includegraphics[width=\linewidth]{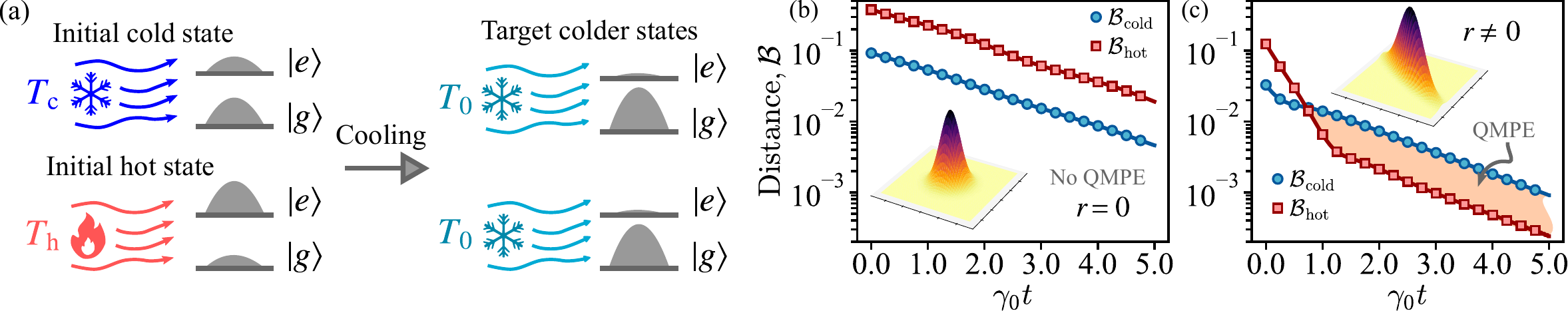}
	\caption{
		(a) Schematic representation of the initial setup: two identical two-level quantum systems are prepared in thermal states at different initial temperatures, \( T_c \) (cold) and \( T_h \) (hot). Each system is then independently cooled down to a common target temperature \( T_0 < T_c \), with population transfer from the excited state \( \ket{e} \) to the ground state \( \ket{g} \) during the cooling process. 
		(b) Time evolution of the systems in contact with a standard (unsqueezed) Gaussian thermal bath (\( r = 0 \)), where the quantum Mpemba effect (QMpE) is absent---that is, the initially hotter system does not relax faster than the colder one. 
		(c) In contrast, the introduction of a non-zero squeezing parameter \( r > 0 \) in the bath allows the engineering of QMpE. A highlighted region marks the regime where the initially hotter system relaxes more quickly than the colder one, a signature of the Mpemba effect emerging due to quantum bath squeezing.
	}
	\label{Fig:Sceme}
\end{figure}

As depicted in Fig.~\SubFig{Fig:Sceme}{a}, a genuine QMpE should be possible if the initial hot and cold states are prepared using the same reservoir at different temperatures, and then the same bath at a colder temperature is used to induce the QMpE. In addition, despite efforts to observe and characterize the QMpE, the specific conditions required to induce this effect and the mechanism behind it remain to be explored in greater detail. In this regard, we address these two questions. Firstly, we conveniently exploit the theory of open quantum systems in the superoperator formalism to derive fundamental conditions that should be satisfied to efficiently achieve QMpE~\cite{Sarandy:05-2}. Such conditions are then used to show that, in the context of reservoir, Hamiltonian, and state preparation engineering, they provide a consistent way of explaining the mechanism behind the emergence of QMpE. With the help of a fidelity-based figure of merit introduced in this work, as shown in Figs.~\SubFig{Fig:Sceme}{b} and~\SubFig{Fig:Sceme}{c}, we identify the emergence of thermally induced QMpE in a two-level atom interacting with squeezed thermal reservoirs. In addition to the application of squeezed states in gravitational-wave detection~\cite{McKenzie:02,Vahlbruch:10,Ligo:11}, quantum metrology~\cite{Anisimov:10,Steinlechner:13}, efficient quantum thermal engines~\cite{Klaers:17,Lindenfels:19,Xiao:23,Niedenzu:18}, among others~\cite{Schnabel:17}, our work establishes the impact of squeezed thermal baths in achieving genuine QMpE.

%%%%%%%%%%%%%%%%%%%%%%%%%%%%%%%%%%%%%%%%%%%%%%%%%%%%%%%%%%%%%%
%%%%%%%%%%%%%%%%%%%%%%%%%%%%%%%%%%%%%%%%%%%%%%%%%%%%%%%%%%%%%%
%%%%%%%%%%%%%%%%%%%%%%%%%%%%%%%%%%%%%%%%%%%%%%%%%%%%%%%%%%%%%%
%%%%%%%%%%%%%%%%%%%%%%%%%%%%%%%%%%%%%%%%%%%%%%%%%%%%%%%%%%%%%%

\section{Sufficient conditions for QMpE}

First of all, let us introduce a function (or quantifier) used as a figure of merit (or witness) of Mpemba's effect. Such a function is defined from the measure of the distance (or infidelity) between two quantum states, introduced in the context of quantum information~\cite{Nielsen:Book}, obtained from the trace distance \(\Bcal_{\mathrm{c},\mathrm{h}}(t) = ||\hat{\rho}_{\mathrm{c},\mathrm{h}}(t) - \hat{\rho}_{0}||_{1}\), with \(||\hat{A}||_{1}=[\mathrm{Tr}(\hat{A}^\dagger\hat{A})]^{1/2}\). \(\hat{\rho}_{\mathrm{c}}(t)\) and \(\hat{\rho}_{\mathrm{h}}(t)\) are solutions of the Lindblad equation for the initial conditions \(\hat{\rho}_{\mathrm{c}}(0)=\hat{\rho}_{\mathrm{c}}\) and \(\hat{\rho}_{\mathrm{h}}(0)=\hat{\rho}_{\mathrm{h}}\), respectively. From this equation, and with the help of Figs.~\SubFig{Fig:Sceme}{b} and~\SubFig{Fig:Sceme}{c}, it is possible to state that whenever the initial states satisfy \(\Bcal_{\mathrm{h}}(0) > \Bcal_{\mathrm{c}}(0)\), Mpemba's effect is observed when \(\Bcal_{\mathrm{h}}(t) < \Bcal_{\mathrm{c}}(t)\), as \(\Bcal\) quantifies the distance to the steady state. Therefore, we define the infidelity-based quantity %~\footnote{It is worth mentioning that in the limit $\tau\rightarrow \infty$ we get $\left\vert \Rcal_{\mathrm{c}}(t) - \Rcal_{\mathrm{h}}(t) \right\vert \rightarrow 0$ and $\left\vert \Bcal_{\mathrm{c}}(t) - \Bcal_{\mathrm{h}}(t) \right\vert \rightarrow 0$, so you can take the limit $\tau\rightarrow \infty$ in Eq.~\eqref{Eq:Qfunction} without loss of generality.}
\begin{equation}
	\Mcal_{\Bcal} = \frac{\Acal_{\Bcal_{\mathrm{c}}>\Bcal_{\mathrm{h}}}}{\Acal_{\Bcal,0}} = \frac{\frac{1}{\tau} \int_{\Bcal_{\mathrm{c}}>\Bcal_{\mathrm{h}}} \left\vert \Bcal_{\mathrm{c}}(t) - \Bcal_{\mathrm{h}}(t) \right\vert dt}{\frac{1}{\tau} \int_{0}^{\tau} \left\vert \Bcal_{\mathrm{c}}(t) - \Bcal_{\mathrm{h}}(t) \right\vert dt} . %~~~ 
	%\Qcal_{\Dcal} = \frac{\Acal_{\Rcal_{\mathrm{c}}>\Rcal_{\mathrm{h}}}}{\Acal_{\Dcal,0}} , 
	\label{Eq:Qfunction}
\end{equation}

The function \(\Acal_{\Bcal_{\mathrm{c}}>\Bcal_{\mathrm{h}}}\) quantifies the (time-averaged) contributions of the fidelity when the hot state gets closer to the target state in comparison to the colder one (as sketched in Fig.~\SubFig{Fig:Sceme}{b}). On the other hand, \(\Acal_{\Bcal,0}\) quantifies the time average of the absolute value of the difference between \(\Bcal_{\mathrm{c}}(t)\) and \(\Bcal_{\mathrm{h}}(t)\) during the total evolution time \(\tau\). To clarify the role of the total evolution time \(\tau\) in the definition of \(\Mcal_{\Bcal}\), notice that the factor \(1/\tau\) appears in both the numerator and the denominator to ensure that \(\Acal_{\Bcal_{\mathrm{c}}>\Bcal_{\mathrm{h}}}\) and \(\Acal_{\Bcal,0}\) are dimensionless quantities, facilitating a proper comparison of the dynamics under different conditions. This normalization does not affect the qualitative interpretation of \(\Mcal_{\Bcal}\), and in the asymptotic limit \(\tau \rightarrow \infty\)---corresponding to the steady state limit---the definition becomes independent of \(\tau\), as both integrals converge. Therefore, the choice of including \(\tau\) in Eq.~\eqref{Eq:Qfunction} is made without loss of generality and is primarily for dimensional consistency.

By construction, \(\Mcal\) has to be smaller than \(1\), since the initial condition of our problem requires us to choose quantum states such that \(\Bcal_{\mathrm{c}}(0) < \Bcal_{\mathrm{h}}(0)\), as depicted in Fig.~\ref{Fig:Sceme}. Also, it is possible to observe that the integral function will capture any contribution of the dynamics in which \(\Acal_{\Bcal_{\mathrm{c}}>\Bcal_{\mathrm{h}}}\), such that \(\Mcal >0\) indicates the emergence of Mpemba's effect within the integration interval considered. Since the absence of Mpemba's effect leads to \(\Mcal = 0\), we can conclude that our parameter obeys \(0 \leq \Mcal < 1\). In this way, the bigger the parameter \(\Mcal\), the stronger Mpemba's effect.

It is worth emphasizing that the Mpemba parameter \(\Mcal_{\Bcal}\), introduced in Eq.~(1), does not correspond directly to a relaxation rate or a specific physical observable. Rather, it is constructed as a mathematical tool to quantify the presence and relative strength of the quantum Mpemba effect. Similar to indicators used in non-Markovian dynamics~\cite{Breuer:09,Rivas:10,Passos:19}, \(\Mcal_{\Bcal}\) captures a global property of the dynamics: in this case, the extent to which the relaxation of the hot state is anomalously faster than that of the cold state over a finite time interval. This provides a practical and scalable approach to detect QMpE in both analytical models and numerical simulations, even when its direct operational interpretation is not accessible.

%For example, the range of integration of the coefficient $\Acal_{\Bcal_{\mathrm{h}}>\Bcal_{\mathrm{c}}}$ is motivated by the the function used as quantifier of Non-Markovianity considered in Refs.~[{\color{red}cite}].

%%%%%%%%%%%%%%%%%%%%%%%%%%%%%%%%%%%%%%%%%%%%%%%%%%%%%%%%%%%%%%%%%%%%%%%%%%%%%%%%%%%%%%%%%%%%%%%%%%%%%%%%%%%%%%%%%%%%%%%%%%%%%%%%%%%%%%%%%%%%%%%%%%%%%%%%%%%%
%%%%%%%%%%%%%%%%%%%%%%%%%%%%%%%%%%%%%%%%%%%%%%%%%%%%%%%%%%%%%%%%%%%%%%%%%%%%%%
%%%%%%%%%%%%%%%%%%%%%%%%%%%% THE QME CONDITIONS %%%%%%%%%%%%%%%%%%%%%%%%%%%%%%
%%%%%%%%%%%%%%%%%%%%%%%%%%%%%%%%%%%%%%%%%%%%%%%%%%%%%%%%%%%%%%%%%%%%%%%%%%%%%%
%%%%%%%%%%%%%%%%%%%%%%%%%%%%%%%%%%%%%%%%%%%%%%%%%%%%%%%%%%%%%%%%%%%%%%%%%%%%%%%%%%%%%%%%%%%%%%%%%%%%%%%%%%%%%%%%%%%%%%%%%%%%%%%%%%%%%%%%%%%%%%%%%%%%%%%%%%%%

Now we introduce new conditions for the emergence of the QMpE through the superoperator formalism for quantum open systems. To this end, our starting point is a time-local Master equation of a quantum system of the form
\begin{equation}
	\frac{d}{dt}\hat{\rho}(t) = \mathfrak{L}[\hat{\rho}(t)] , \label{Eq:LME}
\end{equation}
where \(\hat{\rho}(t)\) is the density matrix describing the system's evolution, and \(\mathfrak{L}[\bullet]\) is the dynamics generator in the system's Hilbert space \(\Hcal\), which encodes the information about the dynamics when the system interacts with external fields and its surroundings, for example. An alternative way to rewrite the above equation, which is convenient to our discussion, is through the \textit{superoperator formalism}~\cite{Sarandy:05-2,Sarandy:05-1,Santos:20d,Alicki:Book07,Petruccione:Book} (see~\ref{ApTitle:Superoperator} for further details). In fact, through this formalism, Eq.~\eqref{Eq:LME} can be rewritten in a ``Schrödinger form" as~\cite{Sarandy:05-2}.

\begin{equation}
	\dket{\dot{\varrho}(t)} = \Lmath \dket{\varrho(t)} , \label{Eq:SLME}
\end{equation}
where \(\dket{\varrho(t)}\) is the \textit{coherence vector} associated with the density matrix \(\hat{\rho}(t)\) with components given by \(\varrho_{n}(t) = \trs{\rho(t) \sigma^{\dagger}_{n} }\), for a basis \(\{\sigma^{\dagger}_{n}\}\) of \(D^2\)-dimensional operators \(\sigma^{\dagger}_{n} \in \Hcal\), with \(D=\dim[\Hcal]\). Similarly, using the same basis, we can define \(\Lmath\) as a \(D^2 \times D^2\) operator with matrix elements \(\Lmath_{kn} = \trs{\sigma_{k}^{\dagger} \mathfrak{L} [ \sigma_{n} ]}\), the \textit{super-operator Lindbladian}. In this way, it is possible to observe that any information about the dynamics generator and the system state is now encoded in \(\Lmath\) and \(\dket{\varrho(t)}\), respectively. As a consequence of this formalism, we can write the solution for the system dynamics as \(\dket{\varrho(t)} = e^{\Lmath t}\dket{\varrho(0)}\). This formalism is convenient for our discussion because we can write the solution \(\dket{\varrho(t)}\) as a linear decomposition of the left- and right-eigenvectors of \(\Lmath\). In fact, without loss of generality, let us assume the set of left- and right-eigenvectors of \(\Lmath\) satisfying \(\Lmath\dket{\Rcal_{n}} = \lambda_{n}\dket{\Rcal_{n}}\) and \(\dbra{\Lcal_{n}}\Lmath = \lambda_{n}\dbra{\Lcal_{n}}\), respectively, then we obtain (see~\ref{ApTitle:Conditions} for further details).

\begin{equation}
	\dket{\varrho(t)} = \sum\nolimits_{n=0}^{D^2-1}e^{\lambda_{n} t} \gamma_{n} \dket{\Rcal_{n}} .
\end{equation}
with coefficients \(\gamma_{n} = \dinterpro{\Lcal_{n}}{\varrho(0)}\). Therefore, we have now shown that the dynamics of the coherence vector is uniquely determined by the initial state and the spectrum of \(\Lmath\). In this way, we can now study the expected behavior of the system in the steady-state, as well as the relevant decay rates of interest for the emergence of the Mpemba effect.

%%%%%%%%%%%%%%%%%%%%%%%%%%%%%%%%%%%%%%%%%%%%%%%%%%%%%%%%%%%%%%
%%%%%%%%%%%%%%%%%%%%%%%%%%%%%%%%%%%%%%%%%%%%%%%%%%%%%%%%%%%%%%
\begin{figure*}
	\centering
	\includegraphics[width=\linewidth]{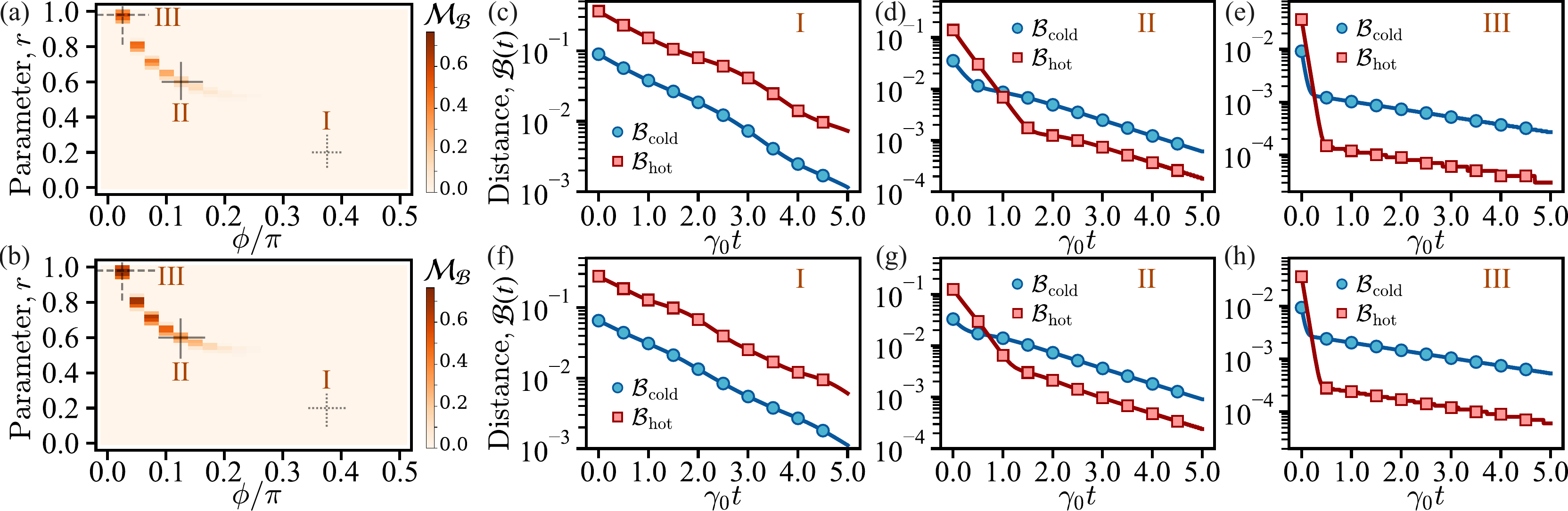}
	\caption{
		(a, b) Mpemba parameter \( \mathcal{M}_{\mathcal{B}} \) as a function of the squeezing parameter \( r \) and the external driving phase \( \varphi \), for two different ratios of system parameters: \( \Omega/\Delta_{0} = 0.5 \) in (a) and \( \Omega/\Delta_{0} = 1.0 \) in (b). This parameter quantifies the occurrence of the quantum Mpemba effect (QMpE), with regions of nonzero \( \mathcal{M}_{\mathcal{B}} \) indicating where the initially hotter system cools faster than the colder one.  
		(c, d, e) Time evolution of the trace distances \( \mathcal{B}_{\mathrm{c}}(t) \) and \( \mathcal{B}_{\mathrm{h}}(t) \), which measure the distinguishability between the instantaneous states (cold and hot initial conditions) and the coldest steady state. These are shown for the specific parameter sets labeled I, II, and III in panel (a), respectively.  
		Similarly, the same analysis is performed in (f, g, h) for the corresponding highlighted points I, II, and III in panel (b).  
		The initial temperatures are set as \( \hbar \omega_0 / (k_B T_c) = 2.0 \) and \( \hbar \omega_0 / (k_B T_h) = 0.1 \) for the cold and hot states, respectively, while the coldest target steady state corresponds to \( \hbar \omega_0 / (k_B T_0) = 100 \). The squeezing phase is fixed at \( \varphi = 0 \).}
	\label{Fig:Mpemba-parameter}
\end{figure*}
%%%%%%%%%%%%%%%%%%%%%%%%%%%%%%%%%%%%%%%%%%%%%%%%%%%%%%%%%%%%%%
%%%%%%%%%%%%%%%%%%%%%%%%%%%%%%%%%%%%%%%%%%%%%%%%%%%%%%%%%%%%%%

First of all, it is worth mentioning that the classical and quantum versions of the Mpemba effect depend only on the initial states of the hot and cold systems, and the common environment used to cool down the systems independently. Within the framework considered here for the QMpE, this means that the hot and cold systems have initial states represented by different coherence vectors, say \(\dket{\varrho_{\mathrm{c}}(0)}\) and \(\dket{\varrho_{\mathrm{h}}(0)}\), respectively, but they are driven through the same super-operator Lindbladian, and therefore we can write down

\begin{equation}
	\dket{\varrho_{\mathrm{c}/\mathrm{h}}(t)} = \sum_{\mathrm{Re}[\lambda_{n}] = 0}e^{\lambda_{n} t} \gamma^{\mathrm{c}/\mathrm{h}}_{n} \dket{\Rcal_{n}} + \sum_{\mathrm{Re}[\lambda_{n}] \neq 0}e^{\lambda_{n} t} \gamma^{\mathrm{c}/\mathrm{h}}_{n} \dket{\Rcal_{n}} . \label{Eq:Evolved}
\end{equation}
where we already separated the sum into two parts associated with all eigenvalues \(\lambda_{n}\) with \(\mathrm{Re}[\lambda_{n}] = 0\), and \(\mathrm{Re}[\lambda_{n}] \neq 0\). The physical intuition behind this distinction relies on the fact that the first term of the above equation may describe the existence of one or multiple steady states, and the second term vanishes for sufficiently large evolution times.

%In fact, as $\mathrm{Re}[\lambda_{n}] \neq 0$ for all $n$ because $\dket{\varrho_{\mathrm{c}/\mathrm{h}}(t)}$ should represent a physical state obeying quantum mechanics postulates, the exponential in the second term vanishes when $t$ is large enough.

In the context of QMpE, the main conclusion of Eq.~\eqref{Eq:Evolved} can be interpreted as follows. By ordering the eigenvalues such that \(\mathrm{Re}[\lambda_{0}] \geq \mathrm{Re}[\lambda_{1}] \geq \cdots \geq \mathrm{Re}[\lambda_{D^2-1}]\), the component of \(\dket{\varrho_{\mathrm{c}/\mathrm{h}}(t)}\) associated with \(\mathrm{Re}[\lambda_{D^2-1}]\) will decay much faster than the component associated with \(\mathrm{Re}[\lambda_{1}]\). This means that a hot state obeying \(|\gamma^{\mathrm{h}}_{1}|\leq|\gamma^{\mathrm{h}}_{2}|\leq \cdots \leq|\gamma^{\mathrm{h}}_{D^2-1}|\) should approach the steady state faster than a cold state with \(|\gamma^{\mathrm{c}}_{1}|\geq|\gamma^{\mathrm{c}}_{2}|\geq \cdots \geq|\gamma^{\mathrm{c}}_{\lambda_{D^2-1}}|\). In addition to these conditions, it is also relevant to observe the relation between the coefficients associated with the eigenvalue with the most negative real part \(\lambda_{D^2-1}\) and the first eigenvalue with a non-zero real part, say \(\lambda_{x}\), such that \(\mathrm{Re}[\lambda_{x}]< 0\). According to Eq.~\eqref{Eq:Evolved}, one expects to be able to observe an enhanced QMpE if the additional relations \(|\gamma^{\mathrm{h}}_{\lambda_{D^2-1}}|\geq |\gamma^{\mathrm{c}}_{\lambda_{D^2-1}}|\) and \(|\gamma^{\mathrm{h}}_{\lambda_{x}}| \leq |\gamma^{\mathrm{c}}_{\lambda_{x}}|\), between the cold and hot initial states are satisfied. 

Notice that, differently from Ref.~\cite{Rylands:23}, we are not providing conditions to \textit{characterize} QMpE (this is done through the parameter \(\Mcal\)); rather, the inequalities stated are expected to be satisfied for any hot and cold states in which the QMpE can be \textit{achieved}. Naturally, obtaining an analytical solution for \(\dket{\varrho_{\mathrm{c}/\mathrm{h}}(t)}\) under arbitrary dynamics is, in general, not feasible. As such, the level of analytical control in our approach only permits us to claim that the derived conditions are \textit{necessary}, but not \textit{sufficient}, for the occurrence of the quantum Mpemba effect (QMpE). Specifically, these conditions must be satisfied for QMpE to emerge within our model, but their fulfillment alone does not ensure the effect will manifest. In contrast, a violation of these conditions is sufficient to rule out the possibility of observing QMpE. This distinction is important: even if the inequalities are satisfied, other factors—such as interference between decay modes or coherence induced by squeezing—may prevent the effect from occurring. Therefore, the proposed conditions should be interpreted as rigorous filters that exclude the QMpE when not met, but do not guarantee its presence when they are. In the following, we illustrate the relevance and applicability of these conditions through a concrete example.

\section{QMpE with squeezed thermal baths}

The system of interest in this work is a single quantum two-level system interacting with a \textit{squeezed thermal bath} at temperature \(T\). As a model, such a system is described by the total Hamiltonian \(\hat{H}_{T} = \hat{H}_{\mathrm{tls}} + \hat{H}_{\mathrm{b}} + \hat{H}_{\mathrm{int}}\). We assume that the two-level system has generic excited and ground states, respectively denoted as \(\ket{e}\) and \(\ket{g}\), such that \(\hat{H}_{\mathrm{tls}} = \hbar \omega_{0} (\ket{e}\bra{e} - \ket{g}\bra{g})\). The thermal bath is described by a system of non-interacting quantum harmonic oscillators as \(\hat{H}_{\mathrm{b}} = \hbar \sum\nolimits_{k} \omega_{k} \hat{b}_{k}^{\dagger}\hat{b}_{k}\), and the bath interacts with the two-level system through the interaction Hamiltonian.

\begin{equation}
	\hat{H}_{\mathrm{int}} = \hbar \sum\nolimits_{k} g_{k}\left[\left(\hat{\sigma}^{+} + \hat{\sigma}^{-}\right)\hat{b}_{k} + \hat{b}_{k}^{\dagger}\left(\hat{\sigma}^{+} + \hat{\sigma}^{-}\right)\right] ,
\end{equation}
where \(\hat{\sigma}^{+} = (\sigma^{-})^\dagger = \ket{g}\bra{e}\), and \(g_{k}\) is the coupling of the atom with the bath mode with wave vector \(k\). The reservoir temperature and squeezing properties are introduced by assuming each mode of the reservoir is initially in the state \(\hat{\rho}_{b,k}(0) = \hat{S}_{k}(r,\varphi) \hat{\rho}_{b,k}^{\mathrm{th}}(0)\hat{S}_{k}^{\dagger}(r,\varphi)\), where \(\hat{\rho}_{b,k}^{\mathrm{th}}(0)\) is the thermal state of the bath modes at temperature \(T\), and \(\hat{S}_{k}(r,\varphi) = \exp\big[ \frac{1}{2} \xi_{k}^{\ast} \hat{b}_{k}^2 - \mathrm{h.c}\big]\) is the squeezing operator, with the squeezing parameters \(r_{k}\) and \(\varphi_{k}\) encoded in \(\xi_{k} = r_{k} e^{i\varphi_{k}}\)~\cite{Manzano:18}. For simplicity, we assume \(\xi_{k} = \xi = r e^{i\varphi}\) for all \(k\). Under these conditions, the dynamics for the two-level system is governed by Eq.~\eqref{Eq:LME}, with the effective Lindblad superoperator \(\mathfrak{L}[\bullet]\) as.

\begin{equation}
	\mathfrak{L}[\bullet] = \frac{1}{i\hbar} [\hat{H}_{\mathrm{tls}},\bullet] + \frac{1}{2} \sum_{n=1}^{2} \left[2\hat{R}_{n}\bullet \hat{R}^{\dagger}_{n} - \{\hat{R}^{\dagger}_{n} \hat{R}_{n}, \bullet  \}\right] , \label{Eq:MasterEq}
\end{equation}
with \(\hat{R}_{1} = \sqrt{\gamma_{0}(N_{\mathrm{th}}+1)} \hat{R}(r,\varphi)\), and \(\hat{R}_{2} = \sqrt{\gamma_{0}N_{\mathrm{th}}} \hat{R}^{\dagger}(r,\varphi)\), where \(\hat{R}(r,\varphi) = \cosh(r)\hat{\sigma}^{-} + e^{i\varphi} \sinh(r)\hat{\sigma}^{+}\), \(\hat{\sigma}^{-} = \hat{\sigma}^{+\dagger} = \ket{g}\bra{e}\), \(\gamma_{0}\) is the spontaneous emission rate, and we assume that the Bose-Einstein statistics applies for the average number of thermal photons \(N_{\mathrm{th}} = 1/(e^{\hbar \omega_{0}/k_{B}T} - 1)\). The above dynamics is valid under the Born-Markov and rotating wave approximations~\cite{Dalton:99,Banerjee:10}. 

This type of reservoir has been efficiently engineered in state-of-the-art nanobeam experiments using squeezed electronic noise~\cite{Klaers:17}, where squeezing parameters $r$ tunable from 0 up to approximately 1 have been demonstrated. For this reason, we will consider this regime of squeezing in our analysis. Comparable levels of squeezing control have also been achieved in optical cavity systems~\cite{Ourjoumtsev:11,Vahlbruch:16} and in parametric up-conversion and down-conversion experiments~\cite{Vahlbruch:08,Heinze:22}, making these platforms promising candidates for the observation of the quantum Mpemba effect.

In this framework, the hot and cold initial states \(\hat{\rho}_{\mathrm{h}}\) and \(\hat{\rho}_{\mathrm{c}}\) are prepared independently as the steady-state of the Eq.~\eqref{Eq:MasterEq} for different temperatures \(T_{\mathrm{h}}\) and \(T_{\mathrm{c}}\), respectively, for a given predefined set of squeezed parameters \(\{r,\varphi\}\). In both cases, the Hamiltonian for the two-level system is the same, and we consider it as being given by the Landau-Zener

\begin{equation}
	\hat{H}_{\mathrm{LZ}} = \hbar \Delta_{0} \hat{\sigma}^{+}\hat{\sigma}^{-} + \hbar \Omega \left(e^{i\phi} \hat{\sigma}^{+} + e^{-i\phi}\hat{\sigma}^{-}\right) .
\end{equation}
which can be implemented by a weak classical oscillating drive (\(\Omega \ll \omega_{0}\)) with frequency \(\omega_{\mathrm{dr}}\) and phase \(\phi\), and detuned by \(\Delta_{0} = \omega_{\mathrm{dr}} - \omega_{0}\) from the atomic transition \(\omega_{0}\). After this stage, each system is placed in contact with the reservoir at very low temperature \(T k_{B}\ll\hbar \omega_{0}\), but keeping the same squeezing parameters used to prepare the hot and cold states, so that the system is driven from their respective initial states according to the Eq.~\eqref{Eq:MasterEq}.

%%%%%%%%%%%%%%%%%%%%%%%%%%%%%%%%%%%%%%%%%%%%%%%%%%%%%%%%%%%%%%
%%%%%%%%%%%%%%%%%%%%%%%%%%%%%%%%%%%%%%%%%%%%%%%%%%%%%%%%%%%%%%
\begin{figure}
	\centering
	\includegraphics[width=0.6\columnwidth]{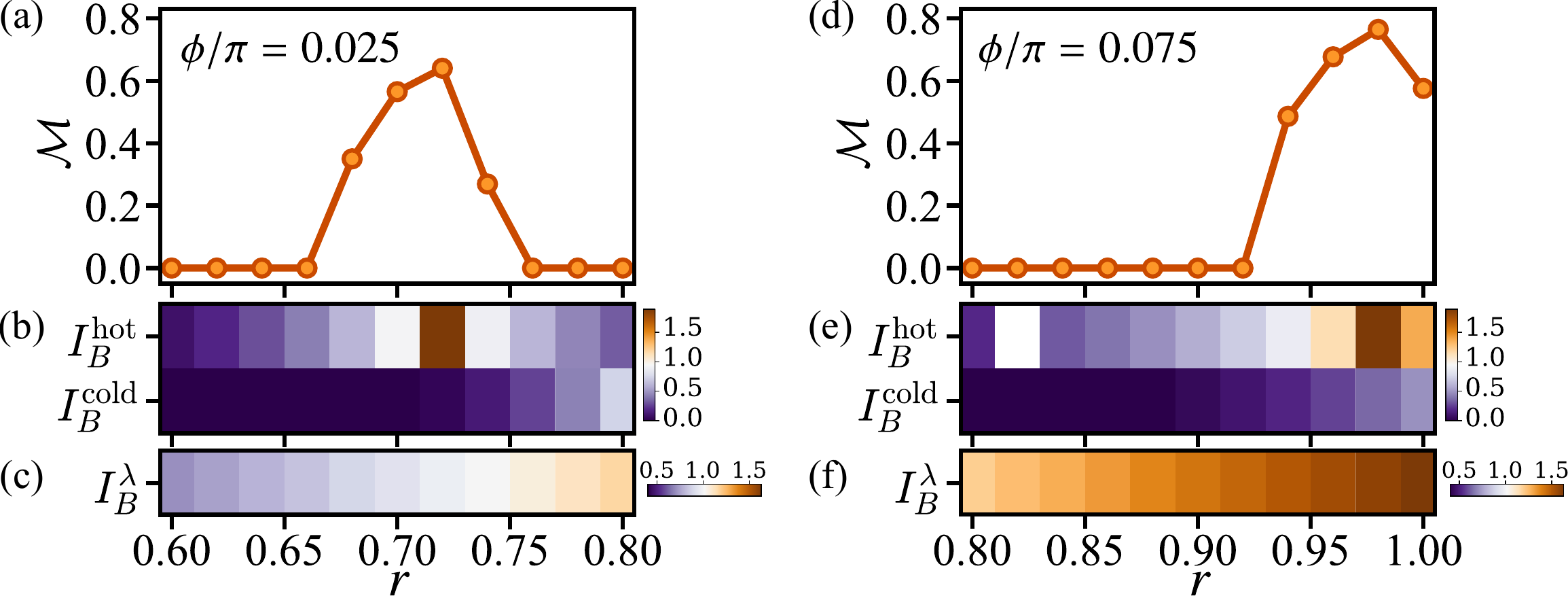}
	\caption{
		Synchronized behavior of key quantities associated with the quantum Mpemba effect (QMpE) as a function of the squeezing parameter \( r \), for two fixed values of the squeezing phase: \( \varphi/\pi = 0.025 \) (left panels) and \( \varphi/\pi = 0.075 \) (right panels).  
		(a, d) Mpemba parameter \( \mathcal{M} \), quantifying the presence of QMpE, shows distinct peaks indicating optimal conditions for the effect as \( r \) is varied.  
		(b, e) Corresponding logarithmic ratios (in bels) of the populations of the coldest reservoir modes for the hot (\( I_{B}^{\mathrm{hot}} \)) and cold (\( I_{B}^{\mathrm{cold}} \)) initial states.  
		(c, f) Logarithmic ratio \( I_{B}^{\lambda} \), representing the spread between the highest and lowest relaxation decay rates (in bels).  
		The panels share a common x-axis (squeezing parameter \( r \)), allowing for direct comparison and revealing a synchronized behavior among \( \mathcal{M} \), \( I_{B}^{\mathrm{hot}} \), \( I_{B}^{\mathrm{cold}} \), and \( I_{B}^{\lambda} \). This synchronization highlights how the emergence of QMpE is linked to changes in both the reservoir mode populations and the decay rate spectrum.}
	\label{Fig:Parameter_cond}
\end{figure}
%%%%%%%%%%%%%%%%%%%%%%%%%%%%%%%%%%%%%%%%%%%%%%%%%%%%%%%%%%%%%%
%%%%%%%%%%%%%%%%%%%%%%%%%%%%%%%%%%%%%%%%%%%%%%%%%%%%%%%%%%%%%%

In Fig.~\ref{Fig:Mpemba-parameter} we show the behavior of the Mpemba parameter \(\Mcal\) as a function of the squeezing parameter \(r\) and the phase \(\varphi\), for two different choices of ratio \(\Omega / \Delta_{0}\), and the behavior of the quantities \(\Bcal_{\mathrm{c}}(t)\) and \(\Bcal_{\mathrm{h}}(t)\) along the evolution for each initial state. In conclusion, it is possible to conclude that the QMpE is not a common effect in the system, as it is only possible to observe it for a small range of parameters. In particular, we highlight the case in which \(r = 0.98\) and \(\phi/\pi = 0.025 \), where the parameter \(\Mcal\) captures strongest Mpemba effect observed in our system, for the regime of parameters shown in Fig.~\SubFig{Fig:Mpemba-parameter}{a} and~\SubFig{Fig:Mpemba-parameter}{b}, respectively. In fact, the Figs.~\SubFig{Fig:Mpemba-parameter}{c},~\SubFig{Fig:Mpemba-parameter}{d} and~\SubFig{Fig:Mpemba-parameter}{e} correspond to the points I, II, and III of Fig.~\SubFig{Fig:Mpemba-parameter}{a}, where one clearly observes that the parameter \(\Mcal\) captures how strong the QMpE is for each case. Furthermore, of particular interest, we highlight the panels in Figs.~\SubFig{Fig:Mpemba-parameter}{f},~\SubFig{Fig:Mpemba-parameter}{g} and~\SubFig{Fig:Mpemba-parameter}{h}, related to the same points in Fig.~\SubFig{Fig:Mpemba-parameter}{b}. By focusing on the cases II and III, we can see that the QMpE gets stronger than in Fig.~\SubFig{Fig:Mpemba-parameter}{a}, and it is mainly related to a faster decay of the hot state in both cases, while the cold states decay slower, leading to a stronger QMpE~\cite{Zhang:24}. 

It means that our analysis reveals that the conditions we propose are more flexible than those required for the genuinely strong quantum Mpemba effect (QMpE) as defined in Ref.\cite{Zhang:24}, where the effect was experimentally demonstrated using a three-level system encoded in the low-lying electronic states of a single $^{40}$Ca$^+$ ion. The strong QMpE necessitates a complete suppression of the population associated with the hot initial state in the subspace corresponding to the slowest-decaying eigenmode. This requirement is consistent with our framework under the constraint $0 = |\gamma^{\mathrm{h}}{1}| \leq |\gamma^{\mathrm{h}}{2}| \leq \cdots \leq |\gamma^{\mathrm{h}}_{D^2-1}|$, which represents an extreme case of our general inequalities. Therefore, while our Mpemba parameter captures an enhanced form of QMpE, it remains a weaker manifestation compared to the stronger version observed in Ref.\cite{Zhang:24}.

While the results aforementioned illustrate the role of the parameter \(\Mcal\) introduced in this work, it does not explain anything about the QMpE. To this end, we now discuss how the conditions established previously allow us to understand the mechanism behind QMpE. To this end, we conveniently define some quantities to support our analysis. In this case, we know the system admits four eigenvalues \(\lambda_{n}\), and then we define the quantity \(I_{B}^{\lambda} = \log_{10} (\mathrm{Re}[\lambda_{3}]/\mathrm{Re}[\lambda_{1}])\), which provides the ratio between the real part of the fastest decay mode \(\lambda_{3}\) and the slowest mode \(\lambda_{1}\) in a Bel scale. This quantity is relevant to our analysis because \(I_{B}^{\lambda}\) quantifies how many orders of magnitude bigger \(\mathrm{Re}[\lambda_{3}]\) is compared with \(\mathrm{Re}[\lambda_{1}]\). Concomitantly, we claim that the QMpE can only be fully understood if we also investigate the relations between the populations \(\gamma_{n}\) for hot and cold states. Therefore, similarly, we also compute the same quantity defined for the ``populations" \(\gamma_{n}\) of the hot and cold states. Of interest to our analysis is the ratio between \(I_{B}^{\mathrm{hot}} = \log_{10} (\mathrm{Re}[\gamma^{\mathrm{hot}}_{3}]/\mathrm{Re}[\gamma^{\mathrm{hot}}_{1}])\) and \(I_{B}^{\mathrm{cold}} = \log_{10} (\mathrm{Re}[\gamma^{\mathrm{cold}}_{3}]/\mathrm{Re}[\gamma^{\mathrm{cold}}_{1}])\). Therefore, we can now explain the emergence of different regimes of Mpemba effect in the system.

In Fig.~\ref{Fig:Mpemba-parameter} we consider the set of parameters corresponding to the points III of the Figs.~\SubFig{Fig:Mpemba-parameter}{a} and~\SubFig{Fig:Mpemba-parameter}{b}. First we show, for each case, how \(\Mcal\) changes with respect to the squeezing parameter \(r\), and we observe that the highest values of \(\Mcal\) in each example are achieved for different choices of \(r\). This is the ideal scenario to test our conditions, so we investigate this behavior taking into account the Fig.~\ref{Fig:Parameter_cond}. By increasing the squeezing parameter \(r\), we create more squeezed thermal photons in the system, which directly affects the ratio \(I_{B}^{\lambda}\). In principle, one could expect emergence of QMpE for any \(r>0.7\), because we have QMpE for \(r = 0.7\) and the decay rate of the mode \(\lambda_{3}\) becomes even faster as \(r\) increases. However, this analysis fails in our system because the slowest decaying mode is significantly populated with respect to the other modes. Therefore we invoke the additional conditions over \(|\gamma^{\mathrm{hot}}_{n}|\) to explain this result.

In Figs.~\SubFig{Fig:Parameter_cond}{a},~\SubFig{Fig:Parameter_cond}{b} and~\SubFig{Fig:Parameter_cond}{c} we observe that even when \( |\gamma_{3,2}| \) is around one order of magnitude bigger than \( |\gamma_{1}| \), the Mpemba parameter detect a moderate QMpE because the population in the mode \( \lambda_{1} \) is big enough to suppress QMpE. However, by combining the fact that \( |\gamma_{3,2}| \gg |\gamma_{1}| > 0 \), as seen in Fig.~\SubFig{Fig:Parameter_cond}{b}, and \( I_{B}^{\lambda} \sim 1 \) (\(\mathrm{Re}[\lambda_{3}] \sim 10 \mathrm{Re}[\lambda_{1}]\)), in Fig.~\SubFig{Fig:Parameter_cond}{c}, we observe the maximum QMpE for the regime of parameters considered is achieved when \( |\gamma_{3,2}| \gg |\gamma_{1}| \). In addition, this effect can be even stronger if we properly choose the Hamiltonian parameters from \( \phi = 0.025\pi \) to \( \phi = 0.075\pi \) (Figs.~\SubFig{Fig:Parameter_cond}{a},~\SubFig{Fig:Parameter_cond}{b} and~\SubFig{Fig:Parameter_cond}{c}), with \( \Mcal \) reaching values close to \( 0.75 \). Similarly, our analysis explains the absence of QMpE even for the squeezed degree providing decay rates satisfying \( \mathrm{Re}[\lambda_{3}] \gg \mathrm{Re}[\lambda_{1}] \). This result is relevant to our discussion because the Hamiltonian parameters can be efficiently controlled through external classical drives. In particular, we notice that in the regime in which QMpE is suppressed, such suppression is related to the fact that the hot state has a significant amount of population in the mode \( \lambda_{1} \). When such a population is sufficiently reduced, we can achieve stronger QMpE with \( \Mcal \approx 0.75 \). It is worth paying attention to the fact that, in all cases considered in Fig.~\ref{Fig:Parameter_cond}, \( I_{B}^{\mathrm{cold}} \) is small compared to \( I_{B}^{\mathrm{hot}} \), which reinforces our claim that the conditions established are \textit{necessary} to achieve stronger QMpE.

\section{Conclusions and prospects}

In conclusion, the results of our work are three-fold. First, we introduced a new parameter to efficiently witness the emergence of QMpE and to quantify how strong the QMpE is. Using this metric as a figure of merit, we developed conditions able to predict scenarios where different regimes of QMpE can be observed. Also, we showed how our results can be applied to explain the experimental observation of strong QMpE reported in Ref.~\cite{Zhang:24}. Finally, we also showed how to achieve different regimes of QMpE through a single two-level system interacting with a squeezed thermal reservoir.

We state that the QMpE superoperator theory and conditions developed in this study go beyond the example considered in this study. For example, our results shed light on the experimental results reported in Ref.~\cite{Zhang:24}. In this work, the authors provided experimental data for the eigenmodes of the Lindbladian of a three-level system encoded in the electronic states of a trapped $^{40}$Ca$^{+}$ ion. By driving such a system through the natural decay relaxation of the system, they investigate the behaviors of the populations of the cold and hot states with respect to the eigenmodes of $\mathfrak{L}[\bullet]$ (slightly different from our definition). As a conclusion, they engineered an initial (pure) strong QMpE state $\ket{sME}$ able to induce a very fast decay of the hot state in comparison with the cold one. As a characteristic of such a state, through the experimental data, they observed that the component of $\ket{sME}$ over the slowest decaying state is numerically zero. This corresponds to the discussions presented for the results shown in Fig.~\ref{Fig:Parameter_cond} of our work.

It is worth mentioning that, as aforementioned, our results can be experimentally reproduced in any experiment where squeezed thermal baths can be created or simulated. In particular, simulations of our findings can be done by digitizing the master equation considered in this work through squeezed Generalized Amplitude Damping channels~\cite{Srikanth:08}. In this regard, our results may be a sufficient, but not necessary, guide to describe both the reservoir and state engineering required to exploit new degrees of QMpE. Also, the understanding of the advantage promoted by squeezed reservoir is still an open question. While our results do not directly demonstrate the underlying physical mechanism, it is plausible that their intrinsically quantum features~\cite{Walls:83,Teich:89}—absent in classical thermal reservoirs—modify the effective decay rates and the Liouvillian structure in a way that favors the suppression of specific decay channels, depending on the squeezing parameters. This could, in principle, create more favorable conditions for the observation of the Mpemba effect.

As final remarks, we note that although our analysis centers on a two-level system in a Markovian regime, the framework we developed is general and applicable to more complex quantum systems. The sufficient condition for the quantum Mpemba effect (QMpE), derived from the Liouvillian spectral structure, is not limited by system dimensionality and it is supported by recent experimental observations in higher-dimensional platforms. Moreover, while additional decoherence mechanisms—such as dephasing or coupling to uncontrolled environmental modes—may shift the regime where the effect is most prominent, they do not preclude its occurrence, as it is governed by population relaxation pathways that remain active even under partial loss of coherence. The device-independent nature of our criteria ensures their robustness across a broad class of dissipative models. Extending this framework to non-Markovian dynamics, particularly involving squeezed or structured reservoirs, presents a promising avenue for further exploration of coherence, dissipation, and thermalization in quantum systems.

\section*{Acknowledgements}

JF would like to thank the Fundação Cearense de Apoio ao Desenvolvimento Científico e Tecnológico (FUNCAP) under the grant PRONEM PNE0112-00085.01.00/16 for financial support, and the Conselho Nacional de Desenvolvimento Científico e Tecnológico (CNPq) under the grant 304485/2023-3. ACS is supported by the Comunidad de Madrid through the program Talento 2024 ``César Nombela", under Grant No 2024-T1/COM-31530 (project SWiQL). ACS also acknowledges the support from the European Union's Horizon 2020 FET-Open project SuperQuLAN (899354) and the Proyecto Sinérgico CAM 2020 Y2020/TCS-6545 (NanoQuCo-CM) from the Comunidad de Madrid.

\appendix

\section{Mathematical framework: Superoperator Formalism}
\label{ApTitle:Superoperator}

In this section we review some fundamental mathematical tools of interest to our development. In particular, we discuss how to obtain: 1) The superoperator representation from the Lindblad and density matrix representation; and 2) The scheme to recover the density matrix representation from the superoperator formalism.

\subsection{The superoperator formalism}

Our discussion is mainly derived from Refs.~\cite{Sarandy:05-1,Sarandy:05-2,Santos:20d,Santos:21c}. First of all, we adopt a \textit{traceless} matrix basis $\{\sigma_{1}, \sigma_{2}, \cdots , \sigma_{D^2-1}\}$ that satisfies $\trs{\sigma_{n}\sigma^{\dagger}_{m}} = D\delta_{nm}$, the density matrix can be written as
\begin{equation}
	\rho = \frac{1}{D} \left[ \1 + \sum_{n=1}^{D^2-1} \varrho_{n} \sigma_{n} \right]  \label{EqEqRhoCoherence}
\end{equation}
with $\varrho_{n} = \trs{\rho \sigma^{\dagger}_{n} }$ and $D$ the dimension of the Hilbert space. As an example, for a single qubit we assume such a basis as given by the set of the Pauli matrices
\begin{equation}
	\sigma_{1} = \sigma_{x} = \left[\begin{array}{cc} 0 & 1 \\ 1 & 0 \end{array}\right] ,
	~~
	\sigma_{2} = \sigma_{y} = \left[\begin{array}{cc} 0 & -i \\ i & 0 \end{array}\right] ,
	~~
	\sigma_{3} = 	\sigma_{z} = \left[\begin{array}{cc} 1 & 0 \\ 0 & -1 \end{array}\right]  ,
\end{equation}
where $\sigma_{0}  = \1$. In this way, the set of Pauli matrices can be used as a matrix basis for a two-level system and we write
\begin{equation}
	\rho = \frac{1}{2} \left[ \1 + \sigma_{x} \varrho_{x} + \sigma_{y} \varrho_{y} + \sigma_{z} \varrho_{z} \right] = \frac{\1 +  \vec{\varrho} \cdot \vec{\sigma}}{2} \mathrm{ , } \label{EqDensiMatrixDecompGen}
\end{equation}
where we define the matrix Pauli vector $\vec{\sigma} = \sigma_{x} \hat{\mathrm{i}} + \sigma_{y} \hat{\mathrm{j}} + \sigma_{z} \hat{\mathrm{k}}$,
with $\{\hat{\mathrm{i}},\hat{\mathrm{j}},\hat{\mathrm{k}}\}$ being the canonical basis of the $\Rmath^3$ Euclidean space, and the \textit{coherence vector} $\vec{\varrho} = \varrho_{x} \hat{\mathrm{i}} + \varrho_{y} \hat{\mathrm{j}} + \varrho_{z} \hat{\mathrm{k}}$ for a two-level system.

Without loss of generality, here we consider the class of evolution where the system dynamics is described by the time-local master equation~\cite{Alicki:Book07,Petruccione:Book}
\begin{equation}
	\dot{\rho}(t) = \Lcalb[\rho(t)] \mathrm{ , } \label{EqEqLind}
\end{equation}
where $\Lcalb[\bullet]$ is the generator of the dynamics of the system interacting with the environment. Thus, if we substitute $\rho(t)$ into Eq.~\eqref{EqEqLind} by using the expanded form of the Eq.~\eqref{EqEqRhoCoherence}, we find the system of differential equations
\begin{equation}
	\dot{\varrho}_{k} (t) = \frac{1}{D} \sum_{n=0}^{D_{-}} \varrho_{i}(t) \trs{\sigma_{k}^{\dagger} \Lcalb [ \sigma_{i} ]} \mathrm{ , } \label{Eqv1}
\end{equation}
where we assume that $\Lcalb [\bullet]$ is a linear superoperator. Note that if we identify the coefficient $\trs{\sigma_{k}^{\dagger} \Lcalb [ \sigma_{i} ]}/D$ in the above equation as an element at $k$-th row and $i$-th column of a $(D^2 \times D^2)$-dimensional matrix $\Lmath$, one can write
\begin{equation}
	\dket{\dot{\rho}(t)} = \Lmath \dket{\rho(t)} \mathrm{ , } \label{ApEq:SuperLindEq}
\end{equation}
where $\dket{\rho(t)}$ is a $D^2$-dimensional vector with components $\varrho_{n}(t) = \trs{\rho(t)\sigma_{n}^{\dagger}}$, $n=0,1,\cdots D_{-}$. 

In particular, for the single-qubit case of interest here $D = 4$, and it means that
\begin{equation}
	\dket{\rho(t)} = [1 ~~ \varrho_{x}(t) ~~ \varrho_{y}(t) ~~ \varrho_{z}(t)] ,
\end{equation}
where all dynamics of the system is mainly ruled by the three last elements of $\dket{\rho(t)}$. This information is important and we will mention it again soon. In an analogous way to the coherence vector $\vec{\varrho}(t)$ of the two-level system, we will call $\dket{\rho(t)}$ the \textit{coherence supervector}.

\subsection{The superoperator spectrum}

As an important remark, this change of formalism has a price to be paid, and it is mainly due to the non-Hermiticity of the superoperator $\Lmath$. Nevertheless, non-Hermitian operators can always be written in the Jordan canonical form, where $\Lmath(t)$ displays a block-diagonal structure $\Lmath_{\mathrm{J}}(t)$ composed by Jordan blocks $J_{n}(t)$ associated with different time-dependent eigenvalues $\lambda_{\alpha}(t)$ of $\Lmath(t)$~\cite{Horn:Book}. When such a Jordan decompoistion is possible, we can write the diagonal form of the superoperator $\Lmath$ as
\begin{equation}
	\Lmath_{\mathrm{J}} = \left[\begin{array}{ccccc}
		J_{k_1}[\lambda_{k_1}] & 0                      & 0        & \cdots & 0 \\
		0                      & J_{k_2}[\lambda_{k_2}] & 0        & \cdots & 0 \\
		\vdots & \ddots & \ddots & \ddots & \vdots \\
		0 & \cdots & 0 & \ddots & 0  \\
		0 & \cdots & \cdots & 0 & J_{k_{N}}[\lambda_{N}] 
	\end{array}\right]_{K \times K} \label{EqEqLindJ} ,
\end{equation}
where $N$ is the number of distinct eigenvalues $\lambda_{\alpha}(t)$ of $\Lmath(t)$ and each block $\Lcal_{k_\alpha}[\lambda_{\alpha}(t)]$ is a $(k_\alpha \times k_\alpha)$-dimensional matrix given as
\begin{equation}
	J_{k}[\lambda_{k}] = \left[\begin{array}{ccccc}
		\lambda_{k} & 1   & 0        & \cdots & 0 \\
		0 &\lambda_{k} & 1 & \cdots & 0 \\
		\vdots & \ddots & \ddots & \ddots & \vdots \\
		0 & \cdots & 0 & \lambda_{k} & 1  \\
		0 & \cdots & \cdots & 0 & \lambda_{k}  
	\end{array}\right]_{k\times k} , \label{EqJordanFormMatrix}
\end{equation}
with $\lambda_{k}$ denoting the eigenvalues of $\Lmath$. Since the Hilbert space of the system has dimension $D$, one finds $k_1 + k_2 + \cdots + k_N = D^2$. In addition, as an immediate consequence of the structure of $\Lmath_{\mathrm{J}}(t)$, one see that $\Lmath(t)$ does not admit the existence of eigenvectors. Instead eigenvectors, we define \textit{right} $\dket{\Rcal_{\alpha}^{n_{\alpha}}}$ and \textit{left} \textit{quasi}-eigenvectors $\dbra{\Lcal_{\alpha}^{n_{\alpha}}}$ of $\Lmath$ associated with the eigenvalue $\lambda_{\alpha}$, satisfying
\begin{eqnarray}\label{EqEqEigenStateL}
	\Lmath\dket{\Rcal_{\alpha}^{n_{\alpha}}} &=& \dket{\Rcal_{\alpha}^{(n_{\alpha}-1)}} + \lambda_{\alpha}\dket{\Rcal_{\alpha}^{n_{\alpha}}} , \\
	\dbra{\Lcal_{\alpha}^{n_{\alpha}}}\Lmath &=& \dbra{\Lcal_{\alpha}^{(n_{\alpha}+1)}} + \dbra{\Lcal_{\alpha}^{n_{\alpha}}}\lambda_{\alpha} .
\end{eqnarray}

The set $\{\dket{\Rcal_{\alpha}^{n_{\alpha}}}\}$ combined with $\{\dbra{\Lcal_{\alpha}^{n_{\alpha}}}\}$ constitutes a basis for the space associated with the operator $\Lmath(t)$ and satisfies the normalization condition $\dinterpro{\Lcal_{\beta}^{m_{\beta}}}{\Rcal_{\alpha}^{n_\alpha}} = \delta_{\beta\alpha} \delta_{m_{\beta}n_{\alpha}}$ and completeness relation
\begin{equation}
	\sum_{\alpha=1}^{N} \sum _{n_{\alpha} = 1}^{N_{\alpha}} \dket{\Rcal_{\alpha}^{n_{\alpha}}}\dbra{\Lcal_{\alpha}^{n_{\alpha}}} = \1_{D^2\times D^2} ,
\end{equation}
where $N$ is the number of Jordan blocks in Eq.~\eqref{EqEqLindJ} and $N_{\alpha}$ is the dimension of the $\alpha$-th Jordan block.

In particular (for the case of interest here), for the cases in which the spectrum of $\lambda_{\alpha}(t)$ is non-degenerate, we can write the uni-dimensional Jordan block decomposition ($N_{\alpha} = 1,~ \forall \alpha$) from the eigenvector equations written as
\begin{eqnarray}
	\Lmath\dket{\Rcal_{\alpha}} =  \lambda_{\alpha}\dket{\Rcal_{\alpha}} , \quad
	\dbra{\Lcal_{\alpha}}\Lmath = \dbra{\Lcal_{\alpha}}\lambda_{\alpha} . \label{EqEqEigenStateUnid}
\end{eqnarray}
and then
\begin{equation}
	\sum_{\alpha=1}^{N} \dket{\Rcal_{\alpha}}\dbra{\Lcal_{\alpha}} = \1_{D^2\times D^2} . \label{ApEq:Identity}
\end{equation}

\section{Superoperator formalism and Mpemba effect}
\label{ApTitle:Conditions}

In this section we explain how the emergence of Mpemba effect observed in the system considered in the main text can be explained using the superoperator formalism. To this end, we move our description to the superoperator formalism, where the system evolution can be properly described by Eq.~\eqref{ApEq:SuperLindEq}. In this formalism, we can write the solution for the Schrödinger-like equation as
\begin{eqnarray}
	\dket{\psi(t)} = e^{\Lmath t} \dket{\psi(0)} 
\end{eqnarray}
where $\Lcalb$ is the superoperator. In particular, using Eq.~\eqref{ApEq:Identity}, it is convenient to write the initial state as a superposition of the right eigenvectors $\dket{\Rcal_{n}}$ of $\Lcalb$ as 
\begin{equation}
	\dket{\psi(0)} = \left[\sum_{n} \dket{\Rcal_{n}}\dbra{\Lcal_{n}}\right]\dket{\psi(0)} = \sum_{n} \Big[\dinterpro{\Lcal_{n}}{\psi(0)}\Big]\dket{\Rcal_{n}}
	= \sum_{n} c_{n}^{L}\dket{\Rcal_{n}},
\end{equation}
with $c_{n}^{L} = \dinterpro{\Lcal_{n}}{\psi(0)}$ the ``projection" of the initial state in the left eigenvector basis of $\Lcalb$. In this way, we can write that
\begin{eqnarray}
	\dket{\psi(t)} = e^{\Lmath t} \dket{\psi(0)}  =  \sum_{n} c_{n}^{L} e^{\Lmath t}  \dket{\Rcal_{n}}
	= \sum_{n} c_{n}^{L} e^{\lambda_{n} t}  \dket{\Rcal_{n}} ,
\end{eqnarray}
where we can explicitly write the steady-state component of the system (associated to $\lambda_{0} = 0$) in this equation as
\begin{equation}
	\dket{\psi(t)} 
	= c_{0}^{L} \dket{\Rcal_{0}} + \sum_{n\neq 0} c_{n}^{L} e^{\lambda_{n} t} \dket{\Rcal_{n}} ,
\end{equation}

Therefore, for a run time large enough we get the steady-state approximation
\begin{equation}
	\dket{\psi(t)} 
	\approx c_{0}^{L} \dket{\Rcal_{0}}  ,
\end{equation}
because $\mathrm{Re}[\lambda_{n}]<0$, and therefore we can observe the emergence of the Mpemba effect only considering the time-dependent term in equation above. In this way, we stablish some sufficient conditions to observe different regimes of Mpemba effect. By ordering $\mathrm{Re}[\lambda_{1}] \leq \mathrm{Re}[\lambda_{2}] \leq \mathrm{Re}[\lambda_{3}]$, we have the conditions to be satisfied for the hot initial state $\dket{\psi_{\mathrm{hot}}(0)}$
\begin{equation}
	|c_{3}^{L,\mathrm{hot}}| < |c_{2}^{L,\mathrm{hot}}| < |c_{1}^{L,\mathrm{hot}}| ,
\end{equation}
with $c_{n}^{L,\mathrm{hot}} = \dinterpro{\Lcal_{n}}{\psi_{\mathrm{hot}}(0)}$, and the cold state satisfies
\begin{equation}
	|c_{3}^{L,\mathrm{cold}}| > |c_{2}^{L,\mathrm{cold}}| > |c_{1}^{L,\mathrm{cold}}| .
\end{equation}
with $c_{n}^{L,\mathrm{cold}} = \dinterpro{\Lcal_{n}}{\psi_{\mathrm{cold}}(0)}$. Any other combination will provide a moderate Mpemba effect.

\section*{References}

%\bibliography{mybib-URL.bib}
%\bibliographystyle{BibStyle}

\end{document}